\documentclass[runningheads]{llncs}
\usepackage[T1]{fontenc}
\usepackage{graphicx}
\usepackage{comment}

\usepackage{cite}
\usepackage{amsmath,amssymb,amsfonts}
\usepackage{algorithmic}
\usepackage{graphicx}
\usepackage{textcomp}
\usepackage{xcolor}
\usepackage{hyperref}

\graphicspath{{images/}} 

\usepackage{color}

\begin{document}
\title{A Mosaic of Perspectives: Understanding Ownership in Software Engineering}

%
%
\author{Tomi Suomi\orcidID{0009-0007-9957-3229} \and Petri Ihantola\orcidID{0000-0003-1197-7266}
\and Tommi Mikkonen\orcidID{0000-0002-8540-9918}
\and Niko Mäkitalo\orcidID{0000-0002-7994-3700}
}
\authorrunning{T. Suomi et al.}
%
\institute{Faculty of Information Technology, University of Jyväskylä, Finland \\
\email{ \{tomi.p.suomi, petri.j.ihantola, tommi.j.mikkonen, niko.k.makitalo\}@jyu.fi}}
\maketitle              
\begin{abstract}
Agile software development relies on self-organized teams, underlining the importance of individual responsibility. How developers take responsibility and build ownership are influenced by external factors such as architecture and development methods. This paper examines the existing literature on ownership in software engineering and in psychology, and argues that a more comprehensive view of ownership in software engineering has a great potential in improving software team's work. Initial positions on the issue are offered for discussion and to lay foundations for further research.
\end{abstract}

\section{Introduction}

Independent and self-organized teams are the cornerstone of agile development. In addition to personal properties, the way developers take responsibility depends on external factors such as architecture or development methods. As an example, microservices have become standard practice, leading to more independently scalable and flexible software architecture at the cost of overall system complexity, where large systems have hundreds or even more microservices, often scattering and obfuscating the lines of ownership \cite{gluckIntroducingDomainOrientedMicroservice2020}. 

Although software ownership has been touched upon in multiple studies \cite{koana2024examining}, the overall understanding of the topic is scarce. The previous research focuses on individual ownership targets, creating various definitions and ways of measuring ownership. The interplay between working practices, organizational structure, architecture, ownership, and various quality attributes is complex. For example, ownership affects various technical quality metrics as well as intangible factors like teamwork \cite{nazir2024understanding} and developer retention \cite{tingting2015impact}. In modern, complex systems, we should better understand the overarching nature of ownership and how to manage it.
 
This paper introduces initial positions on what ownership studies are lacking and how to fill this gap. Section \ref{sec:background} provides background on ownership research, both in the software engineering context and briefly in psychology. Next, Section \ref{sec:argumentation} introduces the research gap and provides six positions based on the analysis of previous research. Finally, Section \ref{sec:researchagenda} derives concrete research questions, discusses potential strategies to answer them, and concludes our work.
\section{Background}
\label{sec:background}

\subsection{Ownership in Software Engineering}

Recent systematic literature review by Koana et al. \cite{koana2024examining} found 28 definitions for ownership, ranging from code ownership to all the way to organizational ownership.  The study focused heavily on \emph{corporeal ownership} side, that refers to development history of an artifact. Moreover, the review divides ownership into three dimensions: What (e.g., code, task, issue, bug, and requirement), Who (e.g, developer, organization, and manager), and How (i.e., dedicated or shared). By using this classification, code was by far the most studied artifact.

\subsubsection{Code ownership}

Bird et al. \cite{bird2011don} studied code ownership as the ratio of commits made by developer against all commits in that particular component. The researchers categorized developers into major and minor contributors, based on whether their total number of commits were more or less than 5\% of total commits, respectively. The study found that high number of minor contributors had correlation with both pre- and post-release failures of both in Windows Vista and Windows 7. Also, top contributors higher proportion of ownership also led to fewer failures, but the correlation was less than when using minor contributors as the variable.

Rahman and Devanbu \cite{rahman2011ownership} on the other hand studied ownership of code at code line level, so that ownership refers to the developer who has authored most of the code. The study also considered developer's specialized (experience in particular file) and general experience. The results suggested that specialized experience is more important in writing bug-free code than general experience. Additionally, faulty code was found to be more likely authored by single developer, contradicting the study by Bird et al. \cite{bird2011don}, where higher number of minor contributors led to more failures. One reason can be the different granularity of metrics, where one study used line level metrics to determine ownership versus commit level, underlining the need for more standard approach on measuring ownership.

Zabarast et al. \cite{zabardast2022ownership} used Ownership and Contribution Alignment Model (OCAM) to understand the alignment of ownership with relation to the actual contributions. Compared to the two previous studies presented above, ticket data was used in addition to Git data to understand ownership and contributions. The authors found that misalignment of ownership (i.e. owning team is not the main contributor to the component) led to faster accumulation of technical debt. The misalignment can happen for example due to module dependencies, where other team needs to do changes in dependent module. Similar pattern was also found by Bird et al. \cite{bird2011don}, where it was observed that minor contributor was often major contributor in dependency package. Such issues can happen with microservice systems for example due to the complex dependencies they might have.

While the studies of code ownership against faults in code are many, the correlation between the two metrics is still unclear. Where Bird et al. \cite{bird2011don} found correlation between number of minor contributors and faults, the replication study by Koana et al. \cite{koana2024examining} didn't see similar effect. And while Koana et al. \cite{koana2024examining} saw increase in bugs with higher number of major contributors, Rahman and Devanbu \cite{rahman2011ownership} noticed that buggy code was more likely authored by single developer. In addition there are studies that found no strong correlation between number of developers and defects, such as the one by Weyuker et al. \cite{weyuker2008too}. The reasons for these inconsistencies can be many, such as different product types, or the context of development such as the methodology used. Deeper understanding of ownership and all of its dimensions should help in eliminating the differences in results, which is why further research is required.

\subsubsection{Psychological ownership}

Psychological ownership in software engineering context was defined as a feeling of ownership toward an owned entity in a project by Koana et al. \cite{koana2024examining}. In study by Sedano et al. \cite{sedano2016practice}, psychological ownership is related to team code ownership (meaning development approach where anyone within team can modify any part of team's code). The authors state that team code ownership is not simply a decree, but rather a feeling. The feeling of team code ownership was found to be supported by understanding the system context, having contributed to the code, perceiving the code quality as high, feeling that the product satisfies user's needs and finally high perceived team cohesion. The study also found multiple risks towards feeling of ownership, such as knowledge silos, increasing code base and team size, and pressure to deliver.

Psychological ownership has also been studied in context of open-source software participant retention by Chung et al. \cite{tingting2015impact}. The authors investigated value and demands-value fit of developers towards open source projects and found that both negatively impact developer turnover in open-source projects. However, higher feelings of psychological ownership were found to moderate the effect on value fit, meaning even if value fit towards the project was low, high feeling of psychological ownership improved developer retention.

Collaboration can also benefit from psychological ownership in agile development. The study by Nazir et al. \cite{nazir2024understanding} investigated both individual and collective (psychological) ownership and suggests that individual ownership can both promote and hinder (e.g. by developer siloing themselves or insisting on solving the issue at hand alone) collective ownership. This aligns well with the discussion on the negative and positive effects of psychological ownership, which are presented in greater detail in Subsection \ref{subsec:psychologicalownership}. The authors found that collective ownership was "turbocharger" in collaboration, meaning collective ownership improved collaboration, which in turn improved collective ownership. The same study states that collective ownership can develop by high perceived level of control over tasks, shared understanding of the tasks and involvement in collaboration. These are well in line with the routes of psychological ownership presented in Fig.~\ref{fig:poSimple}. The authors do note however that collective psychological ownership is not universal concept, and for example organizational culture might affect the feeling of collective psychological ownership.

\subsubsection{Impact of organization}

Looking at the issue of ownership with even wider lens, organizational structure, and therefore the ownership, has also been shown to affect both software failure-proneness as well as test effectiveness and reliability \cite{nagappan2008influence, herzig2014impact}. Herzig and Nagappan \cite{herzig2014impact} studied test effectiveness and reliability, finding that larger organizational subgroups with short communication paths (i.e. test owners are closer to each other in organizational chart) positively correlates with test effectiveness. The study therefore supports the idea that test suites should be owned by individual organizational subgroups. The authors also noted that test suites owned by engineers that have already left the company are less effective.

\subsection{Ownership in Psychology}
\label{subsec:psychologicalownership}

Ownership has been studied vastly in other fields as well, which might provide interesting insights on how to understand ownership in software engineering context. Especially interesting are the ownership studies in psychology, where \emph{psychological ownership} is presented in better detail. Psychological ownership is a feeling of ownership towards a target, or "a cognitive–affective state that characterizes the human condition", held primarily by the individual of this feeling \cite{pierce2003state}.

\begin{figure}[t]
\centerline{\includegraphics[width=0.99\columnwidth]{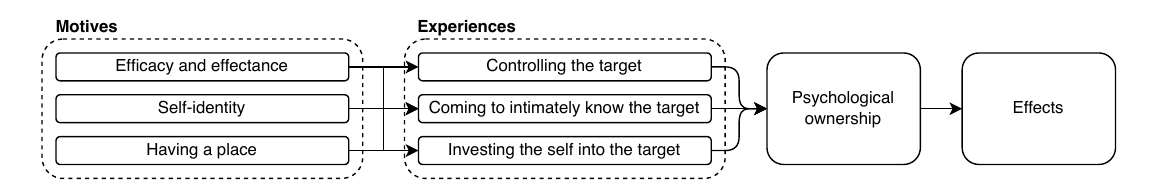}}
\caption{Motives, experiences and effects of psychological ownership \cite{pierce2001toward}.}
\label{fig:poSimple}
\end{figure}

Fig.~\ref{fig:poSimple} represents the motives and experiences that cause psychological ownership as well as consequences resulting from psychological ownership. On the first column of Fig.~\ref{fig:poSimple} we have motives, or roots, of psychological ownership that fulfill basic human motives \cite{pierce2003state}. The first motive efficacy and effectance, meaning as we interact with target, some change happens, leading to feeling of efficacy \cite{white1959motivation}. The second motive is self-identity, referring how objects for example can form and express both private and public identity of a person \cite{wheeler2021objects}. Finally, the third motive is having a place, for example home, which can provide stable refuge, providing security, simulation and identity for the person \cite{porteous1976home}. Having a place has also been described as "belongingness" \cite{avey2009psychological}.

On the second column of Fig.~\ref{fig:poSimple} the major experiences, or routes, from which psychological ownership is born \cite{pierce2001toward} are presented. First there is controlling the target. One way to promote the feeling of control in organizations can be self-managing teams and employee's participation in decision making \cite{liu2012psychological}. Secondly, we have intimately knowing the target. Supporting this, Sedano et al.\cite{sedano2016practice} found that developers in the study felt more ownership towards code when system context was well known to them. Final experience to psychological ownership is investing self into the target. One example is IKEA effect where consumers value products more when they have assembled it themselves \cite{norton2012ikea, sarstedt2017ikea}.

The last column of Fig.~\ref{fig:poSimple} simply represents any potential consequences due to psychological ownership. These can be positive or negative. For example, psychological ownership can lead to higher job satisfaction, commitment and intention to stay within the job \cite{avey2009psychological}. On the negative side, person might refuse to share information or resist change \cite{pierce2001toward}.
\section{Gap in Research}
\label{sec:argumentation}

Based on the related work above, there is little doubt about the importance and industrial impact of ownership. From code quality to test efficiency to technical debt accumulation, understanding ownership can allow companies to exploit this phenomenon. From the psychological ownership side, the increased feeling of responsibility, developer retention, and improved collaboration were just few of the many potential effects. The various known and yet to be discovered benefits of understanding ownership combined with the knowledge of how to foster ownership in practice show promise of significant industry impact. Once completed, the results provide practitioners both the theoretical understanding of ownership as well as how it can be embedded into the existing development processes.

\textbf{Position 1: Ownership research in software engineering has been scattered, lacking full understanding.}
In his paper introducing the chaos model methodology,  Raccoon stated that, "\textit{It seems to me that we have studied each aspect of software development in isolation, not how all aspects fit together.}"  \cite{raccoon1995chaos}. The same appears to be true also for ownership in software engineering: while individual aspects of ownership have been studied, there is no complete model for ownership available that explains how the various pieces work together.  

\textbf{Position 2: Psychological ownership can help understand ownership in software engineering more comprehensively.}
For example, we can consider corporeal ownership and psychological ownership. Corporeal ownership was based on the history of an artifact. But looking at the routes of psychological ownership in Fig.~\ref{fig:poSimple}, it seems unlikely that corporeal ownership exists without psychological ownership. In fact, one paper argued that team code ownership is a feeling \cite{sedano2016practice}, meaning it's related to psychological ownership. Similarly, it could be argued that corporeal ownership is, in fact, psychological ownership. Scenario where developer has created large part of a system but doesn't feel psychological ownership towards it seems hard to imagine. 

\textbf{Position 3: Architectural and methodology choices affect the development and distribution of ownership.}
Continuing on the psychological ownership path, the current assembly line way of software development might have made it more difficult to develop a feeling of psychological ownership towards the entire project. Just like in Ford manufacturing, workers no longer worked on every part of the car, instead they would focus on one or two tasks on an assembly line, which made the work more boring \cite{MovingAssemblyLine} and could arguably have led to loss of feeling of psychological ownership beyond their immediate tasks. It could be argued that software engineering has seen a similar shift towards "assembly line" as well in terms of agile development, where work is iterative and focuses on smaller steps. Similarly microservices architecture breaks the bigger system into smaller unit, affecting the way ownership develops. 

\textbf{Position 4: Time dimension must also be considered in ownership studies and ownership must be transferable.}
While the systematic literature review on ownership in software teams did divide the ownership into components of what, who and how \cite{koana2024examining}, the time component must also be considered. Software engineering teams are dynamic, with people coming and going. Therefore, the ownership model should also be dynamic enough that it doesn't rely on any individual, and even someone with high level of ownership leaving should be manageable without causing unnecessary strain on the development, meaning the ownership must be transferable. 

\textbf{Position 5: To study ownership, we must understand what belongs to software engineers' responsibilities to identify all possible routes to ownership.}
When talking of overarching ownership strategy, it's also important to consider all responsibilities software engineers have. Ayas et al. \cite{michael2024roles} studied the roles, responsibilities, and skills of engineers in microservice era based on job-ads. The authors identified 5 families of responsibilities: software development support \& infrastructure, software product delivery, software process \& team development, professional services delivery and software engineering governance. The implication for ownership study is that there is more to a software engineer's job than simply coding, and ownership studies should not ignore these other responsibilities and how they might affect or be affected by a feeling of psychological ownership. In fact, up to 52\% of developer's workday is used in non-development-heavy activities \cite{meyer2019today}. 

\textbf{Position 6: Ownership must be integrated as part of the software development process.} Once understood more comprehensively, ownership can be used in practice. 

\section{A Research Agenda}
\label{sec:researchagenda}

Inspired by the importance of commitment and self-organization of teams in agile software development, this paper presented the current work done on ownership in software engineering and mirrored research in psychology to software engineering. While in agile methodologies smaller teams can be self-organized, cross-team communication can be challenging in large-scale projects. Although large-scale agile methodologies try to tackle this challenge, we believe ownership can provide interesting pathway to improve agile software development methods at a large scale. It was argued that the ownership research so far has been scattered with no overall understanding of the issue. To guide research further, below are presented research questions to bridge the research gaps discussed.

Based on the position statements above, the following overarching research challenge can be summarized:

\textbf{\emph{How can we embed ownership into software engineering process?}}\\ To decompose this challenge, the following research questions (RQ) are proposed:

\textbf{\emph{RQ1: What is ownership and what does it contain?}} To understand the \emph{targets and dimensions} of ownership, existing literature must be studied in combination with interviews and surveys with software engineering professionals. 

\textbf{\emph{RQ2: How can we measure and influence ownership to understand its benefits?}} To measure and influence ownership, various \emph{instruments} are needed to create standard ways to measure and affect ownership. The existing literature on ownership on software engineering can be compared against known methods of psychological ownership. Also the interviews and surveys from RQ1 can be utilized here.

\textbf{\emph{RQ3: How software development context affects ownership in software teams?}} As software is developed in various contexts (e.g. different methodologies and architectures), their effect on ownership must be understood. The learnings from RQ1 and RQ2 should be studied in practice to understand the fostering of a sense of ownership. Trialing the methods with different contexts (e.g. team using monolith architecture and another team using microservice architecture), nuances of team context can be studied.

The proposed research approach needs action research \cite{staron2020action}, where practitioners are involved directly in the research. The results will then contribute to the body of knowledge by providing practical solutions to understand and benefit from ownership.

\bibliographystyle{splncs04}
\bibliography{references}

\end{document}